# Cross-BCI：A Cross-BCI-Paradigm Classification Model Towards Universal BCI Applications

Gaojie Zhou, Junhua Li, *Senior Member, IEEE*

*Abstract*— Classification models used in brain-computer interface (BCI) are usually designed for a single BCI paradigm. This requires the redevelopment of the model when applying it to a new BCI paradigm, resulting in repeated costs and effort. Moreover, less complex deep learning models are desired for practical usage, as well as for deployment on portable devices. In order to fill the above gaps, we, in this study, proposed a lightweight and unified decoding model for cross-BCI-paradigm classification. The proposed model starts with a tempo-spatial convolution. It is followed by a multi-scale local feature selection module, aiming to extract local features shared across BCI paradigms and generate weighted features. Finally, a multi-dimensional global feature extraction module is designed, in which multi-dimensional global features are extracted from the weighted features and fused with the weighted features to form high-level feature representations associated with BCI paradigms. The results, evaluated on a mixture of three classical BCI paradigms (i.e., MI, SSVEP, and P300), demonstrate that the proposed model achieves 88.39%, 82.36%, 80.01%, and 0.8092 for accuracy, macro-precision, macro-recall, and macro-F1-score, respectively, significantly outperforming the compared models. This study provides a feasible solution for cross-BCI-paradigm classification. It lays a technological foundation for developing a new generation of unified decoding systems, paving the way for low-cost and universal practical applications.

## I. Introduction

BRAIN-computer interface (BCI), a revolutionary human-computer interaction technology, enables users to control external devices or software applications without the involvement of muscles and peripheral nerves by directly decoding neural activity of the brain [1], [2]. The signals acquired through a non-invasive manner could be electroencephalogram (EEG) [3], functional magnetic resonance imaging (fMRI) [4], and functional near-infrared spectroscopy (fNIRS) [5]. Among these signals, EEG is preferable in BCI research and applications due to its lower risk, cost-effectiveness, and practicality [6], [7].

In the area of BCI, there are three classical paradigms: Motor Imagery (MI), Steady-State Visual Evoked Potential (SSVEP),

Gaojie Zhou is with the Laboratory for Brain-Bionic Intelligence and Computational Neuroscience, Wuyi University, Jiangmen 529020, China.

Junhua Li is with the School of Computer Science and Electronic Engineering, University of Essex, CO4 3SQ Colchester, U.K., and also with the Laboratory for Brain-Bionic Intelligence and Computational Neuroscience, Wuyi University, Jiangmen 529020, China (e-mail: junhua.li@essex.ac.uk).

and P300. MI is an active paradigm, which requires participants to actively imagine movements to express their intentions. During MI, event-related desynchronization and synchronization (ERD/ERS) in mu (8–12 Hz) and beta (13–30 Hz) rhythms are obviously induced in the motor cortex without external stimuli and apparent physical movements [8]. When users imagine the movements of a limb, it activates brain regions similar to real physical movements, producing distinct patterns in brain activity that can be reflected by EEG [9], [10]. Therefore, the MI paradigm can be applied in fields such as neurorehabilitation [11], assistive walking with an exoskeleton [12], and prosthetic control [13]. In contrast to the active MI, SSVEP and P300 are typically passive BCI paradigms. They do not rely on the user's voluntary actions, but rather on predictable electrophysiological responses elicited by specific external stimuli presented to the brain [14]. SSVEP refers to the neural oscillations elicited mainly on the visual cortex (for processing visual information) when participants gaze at a visual stimulus flickering at a specific frequency (typically > 4 Hz), which are synchronized with the stimulus frequency and its harmonics [15], [16]. Owing to this characteristic, SSVEP has been widely used in device control [17], [18], [19]. P300 is an event-related potential (ERP) that typically occurs approximately 300 milliseconds after a user observes a rare, task-relevant stimulus (such as a target character flashing or an anomalous sound), resulting in a positive shift in the electroencephalogram signal [20]. Based on this pronounced response to task-relevant events, P300 is commonly used in assistive communication [21], device navigation [22], and language/cognitive rehabilitation [23].

Current decoding models of BCI are usually designed to classify/recognize tasks within a single BCI paradigm. For example, the EEGITNet model, proposed by Salami et al [24], was specifically designed for the MI paradigm and captures tempo-spatial features in EEG signals by dilated causal convolutions. Similarly, the EEG-Inception model, developed by Santamaría-Vázquez et al [25], targets the P300 paradigm, employing a multi-branch convolutional architecture to adaptively decode multi-scale tempo-spatial features of EEG signals. These paradigm-specific models require model redevelopment when applying to a new BCI paradigm, inevitably resulting in additional costs and effort.

Recently, Jiang et al. [26] introduced a large-scale pretrained model, LaBraM, for cross-BCI-paradigm classification in BCI. They utilized eight A800 GPUs to pre-train LaBraM by masked neural code prediction with 2,500 hours of EEG data, and achieved state-of-the-art performance on downstream tasks. Although we acknowledge the advantages of large-scale models in feature learning and generalization, the application of such models in the BCI field faces several challenges. First, the scale of available BCI data remains relatively small compared to domains such as image or video processing. It is still costly to acquire high-quality EEG data [27], making it difficult to meet the huge data requirements of large-scale models. Second, less complex deep learning models are desired for practical usage, as well as deployment on portable devices [28]. Therefore, an effective solution to overcome the current limitations in generalization and practicality of BCI models is to develop models that are lower in complexity, higher in efficiency, and capable of cross-BCI-paradigm classification.

To overcome the challenges mentioned above, this paper presents a novel cross-BCI-paradigm classification model—Cross-BCI. The model first employs a temporal and spatial feature extraction module to form a foundational tempo-spatial joint representation. Then, a multi-scale local feature selection module is designed to extract local features shared across BCI paradigms and generate weighted features. Finally, through the multi-dimensional global feature extraction module, multi-dimensional global features are extracted from the weighted features and fused with the weighted features to form high-level feature representations associated with BCI paradigms. This local-to-global manner enables to extract features shared across BCI paradigms while being capable of generating integrated high-level representations associated with each BCI paradigm, making superior performance in cross-BCI-paradigm classification. The contributions of this study are summarized into the following three points.

1) We have achieved cross-paradigm classification in the context of a small-scale model and limited available data, which is more conducive to applications on portable devices or scenarios requiring low on-chip usage.

2) We develop a multi-dimensional global feature extraction module to form integrated high-level representations associated with each BCI paradigm, thereby enhancing the performance of cross-BCI-paradigm classification.

3) A simple learnable attention mechanism is proposed to weight local features shared across BCI paradigms, which has lower complexity while retaining crucial feature information.

The rest of the paper is organized into the following sections. Section II describes the proposed Cross-BCI model in detail. Section III presents the evaluation settings. Section IV presents the results and discussions. Finally, a brief conclusion is drawn in Section V.

## II. METHODS

The proposed model architecture for cross-BCI-paradigm classification (Cross-BCI) is shown in Fig. 1. It is capable of extracting features shared across BCI paradigms while being capable of generating integrated high-level representations associated with each BCI paradigm. The model architecture is divided into four main modules, which are described in detail as follows.

### A. Temporal and Spatial Feature Extraction Module

In the temporal and spatial feature extraction module, a two-layer convolution is used to learn tempo-spatial joint representation. As shown in Fig. 1, a temporal convolution with the kernel size of $1 \times L_t$ ($L_t$ is a quarter of a second) converts the input EEG into 16 feature maps. $L_t$ is set to a quarter of a second, because it intends to capture the information in frequencies at 4 Hz and above. This setting enables the inclusion of all frequencies of the dominant frequency range of 8–30 Hz for the MI paradigm, as well as the frequencies related to SSVEP stimulus frequencies and their harmonic frequencies (the lowest stimulus frequency is 5.45 Hz in this study). Additionally, the quarter-second window is sufficient to encompass the entire peak, which occurs around 300 milliseconds after the onset of stimuli for the P300 paradigm. The temporal convolution is followed by a spatial convolution.

Spatial convolution employs a convolutional kernel with the size of (C, 1), where C equals the number of EEG channels. This convolution acts as a spatial filter to learn the representation of the interactions between EEG channels [29]. Subsequently, we use the sigmoid linear unit (SiLU) as the activation function for nonlinearity. The tempo-spatial joint representation obtained through this two-layer convolution processing is then fed into the multi-scale local feature selection module.

### B. Multi-Scale Local Feature Selection Module (MSLFS)

We employ the multi-scale local feature selection module to extract local features shared across BCI paradigms and generate weighted features. First, the input feature map $F \in R^{16 \times 1 \times 250}$ is evenly divided into four sub-features $\{F_i\}_{i=1}^{4} \in R^{4 \times 1 \times 250}$ along the filter dimension. The sub-feature $F_1$ is not changed to preserve the tempo-spatial joint representation. The remaining sub-features $F_2$, $F_3$, and $F_4$ are processed through convolutional layers with kernel sizes of $(1 \times 3)$, $(1 \times 5)$, and $(1 \times 7)$, respectively, to extract local features shared across BCI paradigms at different scales. Subsequently, the identical sub-feature $F_1$ is concatenated with the processed sub-features $F_2$, $F_3$, and $F_4$ after multi-scale convolution to form the feature $\mathbf{F}_{cat} \in R^{16 \times 1 \times 250}$. These features are then weighted by the learnable filter attention mechanism, forming the weighted features.

The learnable filter attention mechanism introduces a learnable filter-wise weight vector $w \in R^{16}$, normalised by a softmax function. This vector is used to weight $\mathbf{F}_{cat}$ along the filter dimension, enabling adaptive weighting of features

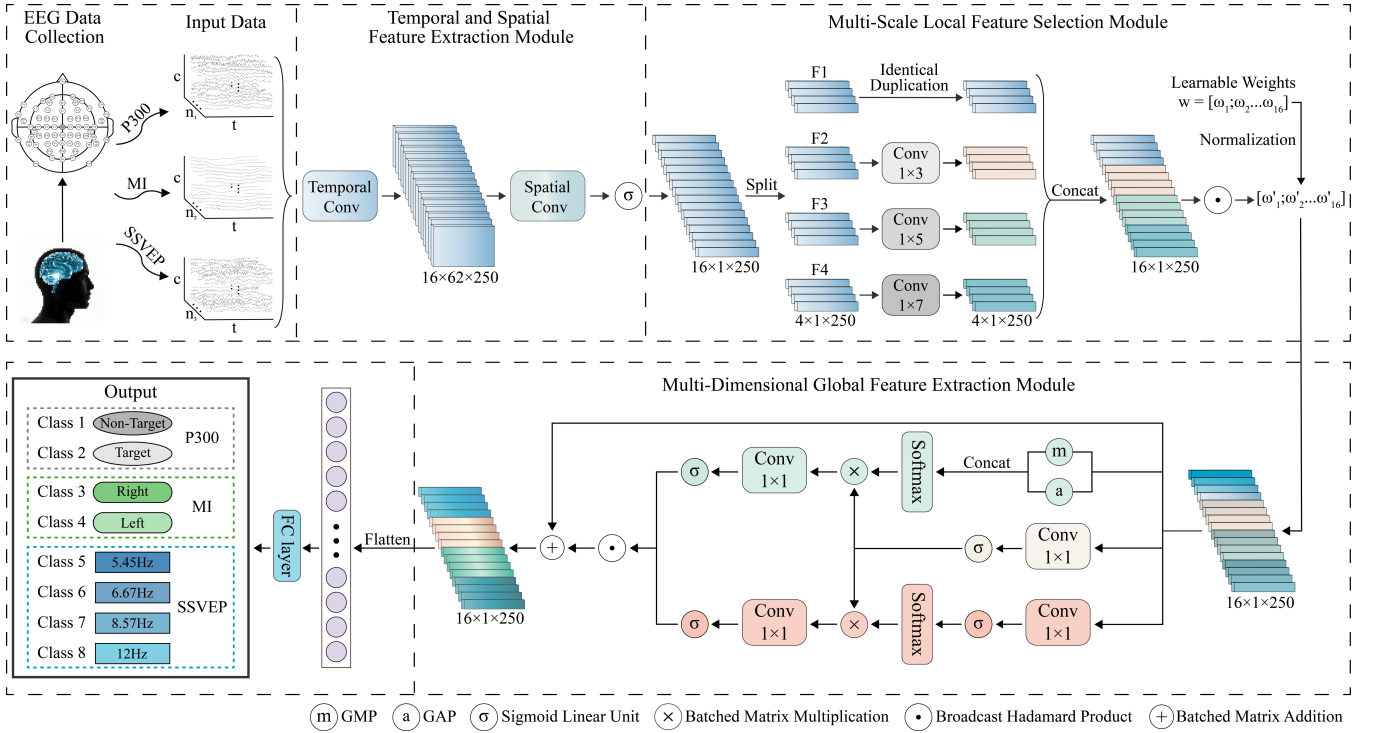

Fig. 1. The proposed model architecture for cross-BCI-paradigm classification (Cross-BCI). It comprises a temporal-spatial feature extraction module, a multi-scale local feature selection module, a multi-dimensional global feature extraction module, and a classification module.

through continuously updated parameters:

$$Z = \text{softmax}(\mathbf{w}) \odot \mathbf{F}_{cat} \quad (1.)$$

where $\odot$ denotes the broadcast Hadamard product, and $Z \in R^{16\times1\times250}$ is the output of the multi-scale local feature selection module.

This design utilises adaptive weighting of features to overcome the limitations of attention mechanisms, such as SENet [30], which commonly reduces the dimensions of extracted features during attention calculation and negatively affects prediction accuracy. In addition, this design effectively emphasise informative features while reducing computational complexity.

### C. Multi-Dimensional Global Feature Extraction Module (MDGFE)

After extracting local features shared across BCI paradigms and generating weighted features through the multi-scale local feature selection module, we designed a multi-dimensional global feature extraction module, in which multi-dimensional global features are extracted from the weighted features and fused with the weighted features to form high-level feature representations associated with BCI paradigms. This module consists of three parallel branches. The first branch is a filter attention branch based on global pooling, which applies global average pooling (GAP) and global max pooling (GMP) to the feature map $Z$ to integrate global information:

$$B_1 = \text{Concat}(\text{GAP}(Z); \text{GMP}(Z)) \quad (2.)$$

where $B_1 \in R^{2\times1\times16}$ denotes the output of the first branch.

The second branch is a linear mapping branch, which applies a convolutional layer with a kernel size of $(1 \times 1)$ to the feature map $Z$ for linear transformation, aiming to construct a feature vector corresponding to the value vector in the self-attention mechanism [31]:

$$B_2 = V(Z) \quad (3.)$$

where V refers to the convolutional layer that performs a linear transformation on the feature map $Z$, and $B_2 \in R^{1\times16\times250}$ denotes the output of the second branch.

The third branch is the spatial attention branch, which uses a convolutional layer with a kernel size of $(1 \times 1)$ to generate an attention score matrix similar to the multiplication of the query and key in the self-attention mechanism, thereby reducing the number of model parameters:

$$B_3 = K(Z) \quad (4.)$$

where K refers to the convolutional layer that generates the attention score matrix, and $B_3 \in R^{1\times250\times1}$ denotes the output of the third branch.

Subsequently, the outputs of the first and third branches are normalised using the softmax function and then respectively multiplied by the output of the second branch, resulting in global contextual information across filters and spatial dimensions:

$$A_1 = \text{softmax}(B_1) \otimes B_2 \quad (5.)$$
$$A_2 = \text{softmax}(B_3) \otimes B_2 \quad (6.)$$

where $A_1 \in R^{2\times1\times250}$ and $A_2 \in R^{16\times1\times1}$ represent the global contextual information across filters and spatial dimensions, respectively.

Finally, A1 and A2 are adjusted using convolutions, which is followed by an operation of broadcast Hadamard product to generate multi-dimensional global features that are fused with the weighted features to form high-level feature representations associated with BCI paradigms:

$$C = Z \oplus (M_1(A_1) \otimes M_2(A_2)) \quad (7.)$$

where $C \in R^{16 \times 1 \times 250}$ is the output of the multi-dimensional global feature extraction module, and M1 and M2 are convolutional layers used for adjustment. The high-level feature representations are then fed into the classification module.

### D. Classification Module

The classification module consists of a flatten operation and a fully connected layer. The output of the module is the final classification results.

## III. EVALUATION SETTINGS

### A. Data Description

OpenBMI is one of the popular datasets [32], which includes three typical BCI paradigms (MI, P300, and SSVEP) we discuss in this study. A total of 54 human participants completed two sessions on separate days. Each session consisted of P300, MI, and SSVEP tasks performed sequentially. Each paradigm comprises two components: an offline phase (without decoding) and an online phase (involving real-time decoding and feedback) part. In the MI paradigm, participants sequentially underwent a three-second preparation, followed by four seconds of left or right hand movement imagery, and a rest period lasting six seconds with a random fluctuation of ±1.5 seconds. The four-second EEG data corresponding to the motor imagery task phase were extracted and segmented into four one-second samples, resulting in a total of 1,600 MI samples for each participant. In the P300 paradigm, a classic 6×6 matrix speller (A to Z, 1 to 9, and _) was used. The stimulus-time interval was set to 80 ms, and the inter-stimulus interval (ISI) to 135 ms. Each target character underwent five rounds of stimulation sequences, each of which consists of 12 row/column flashes. A one-second segment of EEG data following each flash was extracted and considered as a sample. There are a total of 8,280 samples of P300 for each participant. In the SSVEP paradigm, flicking stimuli at 5.45, 6.67, 8.57, and 12 Hz were used. Each SSVEP stimulus was presented for four seconds with an ISI of six seconds. Each target frequency was presented 25 times. We extracted four seconds of EEG data corresponding to the stimulus phase and divided them into four one-second samples. There are a total of 1,600 samples of SSVEP for each participant. In total, there are eight classification categories across the three paradigms (MI: 2 classes; SSVEP: 4 classes; P300: 2 classes), and each participant obtained 11,480 EEG samples. EEG signals were recorded using 62 Ag/AgCl electrodes at a sampling rate of 1,000 Hz. The amplifier used in the experiment was a BrainAmp (Brain Products; Munich, Germany). The impedances of the EEG electrodes were maintained below 10 kΩ during the entire experiment. Further experimental details can be found in the literature [32].

In terms of preprocessing, we only performed the minimal necessary processing. First, a second-order Butterworth band-pass filter (0.5~80 Hz) was applied to remove low-frequency drifts and high-frequency noise. Subsequently, a 60 Hz notch filter was used to suppress power line interference. Finally, all EEG samples were downsampled to 250 Hz and normalized.

### B. Experiment Settings

Due to the non-stationary nature of EEG signals, they exhibit significant dynamic variations across different times or locations [33]. Therefore, to evaluate the generalization ability of the proposed cross-BCI-paradigm model, five-fold cross-validation (CV) was employed. Specifically, in each fold of CV, all samples from a single participant were randomly divided into a training set (80%) and a testing set (20%). Model training was conducted on the training set. After 50 epochs, classification accuracy on the training set was monitored for each epoch. Training would stop if the training accuracy was not further improved for 10 consecutive epochs or if the number of epochs exceeded 1000. Finally, the model that achieved the highest classification accuracy on the training set was selected for model evaluation on the testing set. Accuracy (Acc), macro-precision (Mac-Pre), macro-recall (Mac-Rec), and macro-F1-score (Mac-F1) are used as evaluation metrics, which are formularized as

$$\text{Acc} = \frac{TP + TN}{TP + TN + FP + FN} \quad (8.)$$

$$\text{Mac-Pre} = \frac{1}{N} \sum_{i=1}^{N} \frac{TP_i}{TP_i + FP_i} \quad (9.)$$

$$\text{Mac-Rec} = \frac{1}{N} \sum_{i=1}^{N} \frac{TP_i}{TP_i + FN_i} \quad (10.)$$

$$\text{Mac-F1} = \frac{1}{N} \sum_{i=1}^{N} \frac{2 \cdot TP_i}{2 \cdot TP_i + FP_i + FN_i} \quad (11.)$$

where TP is the true positive, TN is the true negative, FP is the false positive, FN is the false negative, and the subscript indicates the i-th class.

### C. Compared Models

The following models were selected for comparison because they represented both classical and cutting-edge approaches in EEG decoding.

EEGNet [34]: EEGNet uses depth-wise convolution to learn spatial filters under each temporal filter and fuses different spatial filters through point-wise convolution. EEGNet is a compact convolutional neural network specifically designed for EEG-based BCIs.

DeepConvNet [35]: DeepConvNet comprises five convolutional blocks: an initial convolution-pooling layer followed by three standard convolution-max-pooling blocks, and a final dense softmax classification layer.

EEG-Inception [25]: Designed as a classification tool for P300-based spellers, EEG-Inception is a CNN architecture that comprises two inception modules and an output module.

EEGITNet [24]: EEGITNet leverages inception modules and dilated causal convolutions to effectively extract spectral, spatial, and temporal features.

### D. Training Details

In our study, we used log cross-entropy loss with the RAdam optimizer. Both learning rate and weight decay were set to 0.001. Batch size was set to 32. To fit the OpenBMI dataset, we had to slightly adjust temporal convolution kernels of the compared models. For EEG-Inception and EEGNet, we used input data with a sampling frequency of 250 Hz. In contrast, EEGNet and EEG-Inception were initially designed for data with a sampling frequency of 128 Hz, and both use one-second segments as input. Therefore, we adjust the length of time convolution kernels in both architectures by a factor of 2 to approximate the 250Hz sampling rate, consistent with previous work [36]. DeepConvNet originally uses two-second segment as input, but the segment length is one second in our study. Therefore, the size of temporal convolution kernel was adjusted to align with one-second segment. Except the above modifications, all other settings of the compared models, including hyperparameters, optimizer, and loss functions, were kept the same to that described in their original papers. All models were evaluated on the same desktop computer equipped with a GeForce GTX 2080Ti GPU with 11 GB of memory.

## IV. RESULTS AND DISCUSSIONS

### A. Comparison Results

We conducted a comprehensive comparison between the proposed Cross-BCI model and the compared models across three aspects to validate the effectiveness of the proposed model in cross-BCI-paradigm classification.

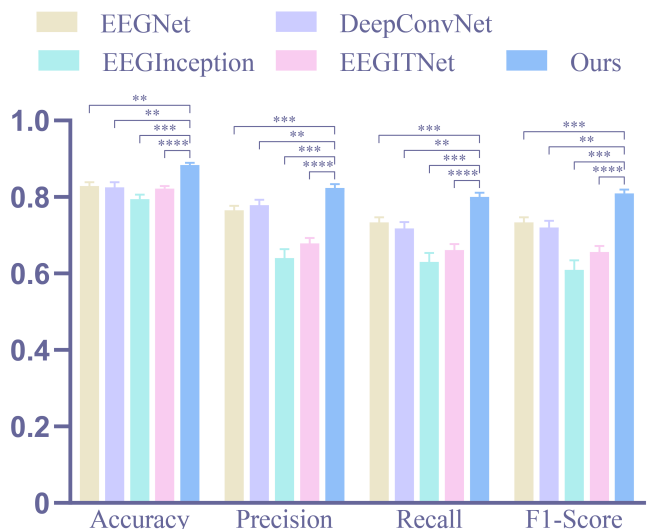

Fig. 2. Cross-BCI-paradigm classification results of the proposed model and the compared models. Whiskers indicate standard errors. Asterisks indicate the level of statistical significance (* p<0.05, ** p<$10^{-5}$, *** p<$10^{-10}$, **** p<$10^{-20}$).

1) Comparison of Average Classification Performance: As shown in Fig. 2, the Cross-BCI model achieves the best performance in terms of accuracy, macro-precision, macro-recall, and macro-F1-score, which are 88.39%, 82.36%, 80.01%, and 0.8092, respectively. Compared to EEGNet, DeepConvNet, EEGInception, and EEGITNet, Cross-BCI model demonstrated improvements in accuracy of 5.50%, 5.89%, 8.90%, and 6.21%; demonstrated improvements in macro-precision of 5.80%, 4.48%, 18.34%, and 14.52%; demonstrated improvements in macro-recall of 6.61%, 8.25%, 17.01%, and 13.87%; and demonstrated improvements in macro-F1-score of 0.0756, 0.0891, 0.1994, and 0.1527. The standard errors of Cross-BCI model are minimal in all four metrics among those models, indicating that the proposed model is more reliable in the cross-BCI-paradigm classification. In addition, a paired t-test was conducted to assess whether the performance differences between the proposed Cross-BCI model and the compared models are statistically significant. The results show that for all metrics, the p-values are less than $10^{-5}$, indicating that the performance differences between Cross-BCI and the compared models are statistically significant.

2) Comparison of Individual Classification Performance: Given the significant inter-participant variability in EEG signals [37], we explored the comparisons separately for each participant. As shown in Fig. 3, Cross-BCI model outperforms all compared models in terms of accuracy for every participant. In terms of macro-F1-score, Cross-BCI outperforms the compared models for 53 out of 54 participants. These results indicate that Cross-BCI highly consistently outperforms the compared models in cross-BCI-paradigm classification, demonstrating greater robustness to inter-participant variability.

3) Comparison of Category Classification Performance: We compared the models in the form of confusion matrix to see the classification performance of each category. The results are shown in Fig. 4. Cross-BCI achieves the highest performance for all categories except the category of P300 target. For the category of P300 target, Cross-BCI was second, worse than EEGITNet, but better than all other compared models. This result indicates that Cross-BCI model is more effective to discern the differences between categories among different BCI paradigms.

### B. Ablation

To demonstrate the role of the key modules in Cross-BCI model, we conducted an ablation study. The results are shown in Table I. It can be observed that the removal of both the MSLFS module and the MDGFE module led to the larger reduction in the performance (decreased by 4.75%) compared to the removel of a single module (decreased by 4.23% when removing the MDGFE module, and decreased by 1.23% when removing the MSLFS module). These results collectively suggest that the key modules within Cross-BCI play a vital role, and their synergistic operation largely enhances the performance of cross-BCI-paradigm classification.

We also explored the distributions of the features extracted by the Cross-BCI model with or without the key modules. The extracted high-dimensional features before the fully-connected layer were projected into a two-dimensional space by t-SNE, and were then visualized in Fig. 5. The visualization shows that the features are more separable among categories when both the MSLFS module and the MDGFE module are included in Cross-BCI model. This finding is highly consistent with the results of the ablation study, further demonstrating that the simultaneous use of the two key modules in the Cross-BCI model enables the learning of the most discriminative features, thereby significantly enhancing cross-BCI-paradigm classification performance.

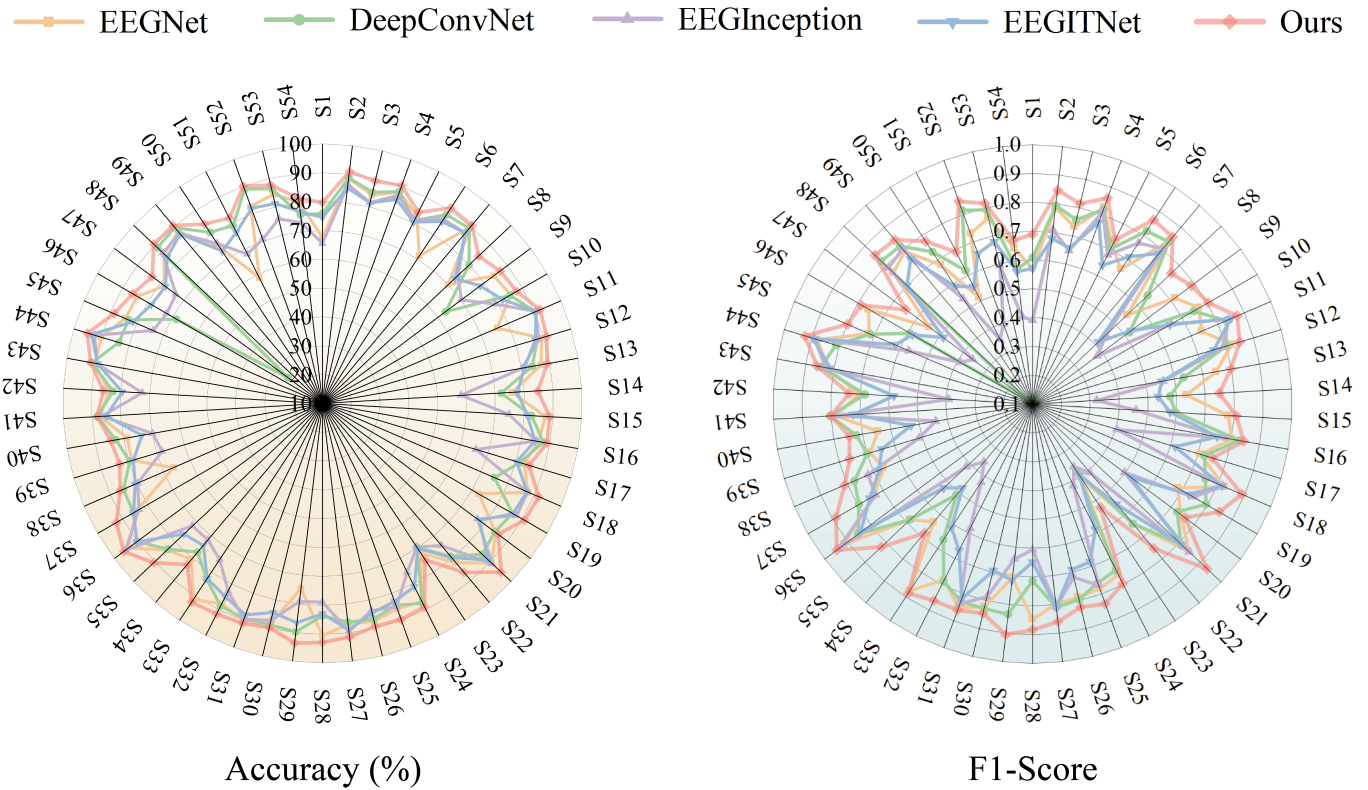

Fig. 3. The results of cross-BCI-paradigm classification for individual participants.

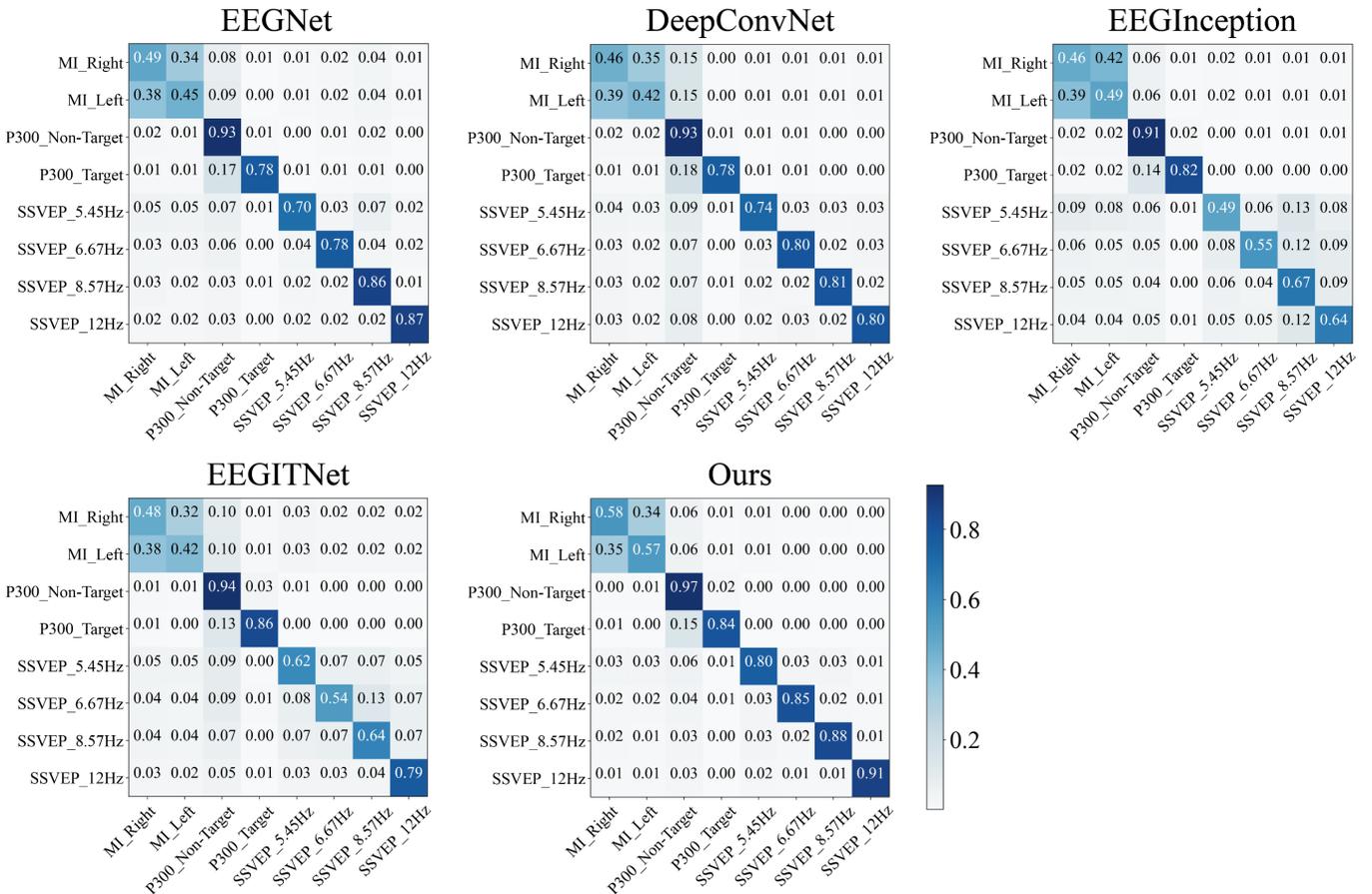

Fig. 4. The confusion matrices of the proposed model and the compared models.

TABLE I
RESULTS OF ABLATION STUDIES PERFORMED ON THE OPENBMI DATASET, WITH BEST RESULTS MARKED IN BOLD

| Method | Acc | STD | Mac-Pre | STD | Mac-Rec | STD | Mac-F1 | STD |
|---|---|---|---|---|---|---|---|---|
| w/o MSLFS&MDGFE | 83.64% | 5.52% | 75.25% | 9.63% | 72.24% | 10.52% | 0.7345 | 10.21% |
| w/o MDGFE | 84.16% | 5.17% | 75.38% | 9.45% | 73.62% | 9.67% | 0.7430 | 9.60% |
| w/o MSLFS | 87.16% | 4.25% | 80.41% | 7.76% | 77.44% | 8.74% | 0.7867 | 8.36% |
| Ours | **88.39%** | **4.12%** | **82.36%** | **7.26%** | **80.01%** | **8.07%** | **0.8092** | **7.76%** |

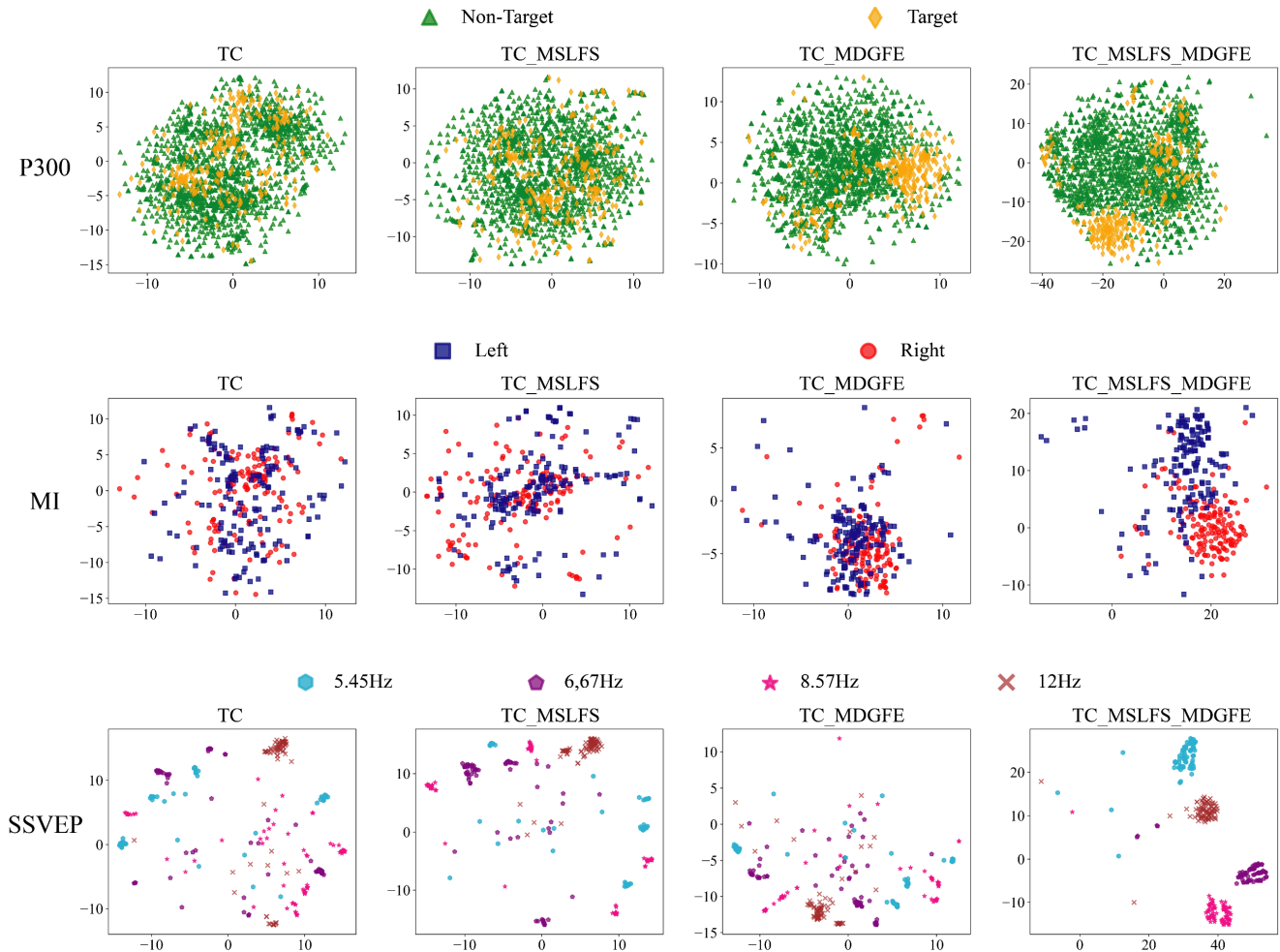

Fig. 5. Visualization of the extracted features by the Cross-BCI model for an example participant (Participant 36).

TABLE II
COMPARISON RESULTS OF DIFFERENT FILTER ATTENTION MECHANISMS, WITH BEST PERFORMANCE MARKED IN BOLD

| Method | Acc | STD | Mac-Pre | STD | Mac-Rec | STD | Mac-F1 | STD |
|---|---|---|---|---|---|---|---|---|
| SENet | 87.43% | 4.15% | 80.81% | 7.48% | 78.16% | 8.37% | 0.7925 | 8.02% |
| ECANet | 87.43% | 4.16% | 80.87% | 7.52% | 77.96% | 8.59% | 0.7916 | 8.18% |
| Ours | **88.39%** | **4.12%** | **82.36%** | **7.26%** | **80.01%** | **8.07%** | **0.8092** | **7.76%** |

*C. Comparative Analysis of Different Filter Attention Mechanisms*

In the MSLFS module, we compared the proposed filter attention mechanism with two commonly used filter attention mechanisms (SENet [30] and ECANet [38]). SENet uses a network of two fully-connected layers to learn filter attention, in which the first fully-connected layer performs feature dimensionality reduction, and the second fully-connected layer restores the feature dimension back to its original size. This suffers from the drawbacks of information loss caused by dimensionality reduction and a relatively large number of parameters. As listed in Table II, its performance on all four cross-BCI-paradigm classification evaluation metrics is lower than that of our filter attention mechanism. ECANet employs one-dimensional convolution to enable local cross-filter interaction, achieving slightly better classification performance compared to SENet. It also has the drawback of a relatively large number of parameters. In contrast, our proposed learnable filter attention mechanism is a simple and efficient mechanism that introduces only a learnable filter-wise weight vector, which directly weights on the features. Compared to SENet and ECANet, it achieved the best performance (see the comparison details in Table II).

*D. Online Simulation*

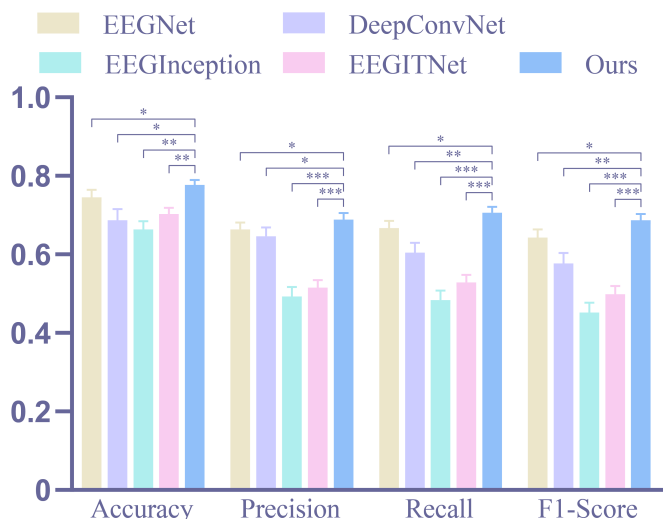

Fig. 6. Results of simulated real-time classification. Whiskers indicate standard errors. Asterisks indicate the level of statistical significance (* $p<0.05$, ** $p<10^{-5}$, *** $p<10^{-10}$, **** $p<10^{-20}$).

We simulated the practical application of BCI classification. Specifically, the offline phase data (without decoding) from OpenBMI dataset were used to train the models. The online phase data (involving real-time decoding and feedback) were used to evaluate the models in the same manner as during real-time testing in data acquisition. Except for the sample length, our study adopts 1-second data segments as the sample length. The parameter settings for the models are the same as described in Section III. According to the comparison results (see Fig. 6), Cross-BCI model significantly outperformed the other models across all four evaluation metrics, as well as possessing the lowest standard error. Furthermore, the paired t-test confirms that Cross-BCI model's performance improvement is statistically significant, with p-values less than 0.05. These results demonstrate that the Cross-BCI model exhibits excellent classification performance and reliability in the real-time classification scenario, thereby providing theoretical support for the practical deployment of cross-BCI-paradigm systems.

## V. Conclusion

In this paper, we presented a novel deep learning model, named Cross-BCI, aiming to cross-BCI-paradigm classification without retraining and retuning. The proposed model was compared to the state-of-the-art models under four evaluation metrics on a mixture of three classical BCI paradigms (i.e., MI, SSVEP, and P300). The results show that the proposed model achieves 88.39%, 82.36%, 80.01%, and 0.8092 in terms of accuracy, macro-precision, macro-recall, and macro-F1-score, respectively, significantly outperforming the compared models. We further investigated the effect of key modules in our proposed model on classification performance. The ablation study and feature visualization suggest that the key modules play a positive role in the improvement of the cross-BCI-paradigm classification performance. Moreover, we simulated the real-time classification scenario to evaluate the online classification performance. The simulation demonstrates that the Cross-BCI model still exhibits excellent classification performance and reliability, significantly outperforming the compared models. In the future, we plan to apply it for the development of a low-cost, general-purpose, and new unified BCI decoding system